\documentclass[preprint,pre,aps,showpacs]{revtex4}
\usepackage{graphicx}
\begin{document}

\title{Isoscaling, Symmetry Energy and Thermodynamic Models} 

\author{G. Chaudhuri$^1$, S. Das Gupta$^1$, and M. Mocko$^2$}

\affiliation{$^1$Physics Department, McGill University, 
Montr{\'e}al, Canada H3A 2T8}

\affiliation{$^2$Los Alamos National Laboratory, P. O. Box 1663, Los Alamos, NM 87544, USA}

\date{\today}

\begin{abstract}

The isoscaling parameter usually denoted by $\alpha$ depends
upon both the symmetry energy coefficient and the isotopic
contents of the dissociating systems.  We compute $\alpha$
in theoretical models: first in a simple mean field model
and then in thermodynamic models using both grand canonical
and canonical ensembles.  For finite systems the canonical
ensemble is much more appropriate.  The model values of $\alpha$
are compared with a much used standard formula.  Next we turn
to cases where in experiments, there are significant deviations
from isoscaling.  We show that in such cases,
although the grand canonical model fails, the canonical model is capable 
of explaining the data.

\end{abstract}


\maketitle

\section{Introduction}

For central collisions of Sn on Sn ($^{112}$Sn+$^{112}$Sn, $^{124}$Sn+
$^{112}$Sn and $^{124}$Sn+$^{124}$Sn) a well-known result is that the ratio
of isotope yields from two different reactions, 1 and 2, $R_{21}(N,Z)=
Y_2(N,Z)/Y_1(N,Z)$ exhibits an exponential relationship as a function of
the isotope neutron number $N$ and proton number $Z$ 
\cite{Xu,Tsang1,Tsang2}:
\begin{eqnarray}
R_{21}(N,Z)=Y_2(N,Z)/Y_1(N,Z)=C\exp(\alpha N+\beta Z)
\end{eqnarray}
This is called isoscaling.  Note that for Sn on Sn central collisions
the fragmenting system is 
rather large.  When the fragmenting system is significantly smaller,
the above equation is only approximate \cite{Mocko1}.  
We will confine ourselves to large systems till we come to sections
VII and VIII.

Much effort has gone into trying to relate $\alpha$
to the symmetry energy term that occurs in liquid drop binding energy
formula.  In its simplest version the symmetry
energy term is given by $C_s(N-Z)^2/A$.
It is reasonable to guess that the ratio $R_{21}(N,Z)$ should
predominantly depend on $N_0,Z_0$ or equivalently on $Z_0,A_0$ of
the fragmenting systems and also on the value of $C_s$.  An 
approximate functional relationship that can be deduced from models is:
$\alpha\approx 4\frac{C_s}{T}((Z_0(1)/A_0(1))^2-(Z_0(2)/A_0(2))^2)$
where $Z_0(1)/A_0(1)$ refers to the disassociating system in reaction 1,
$Z_0(2)/A_0(2)$ refers to that in reaction 2 and $T$ is the characteristic
temperature in the two reactions.  It is this approximate equality 
that we examine in this work, first in a mean-field model (section III)
and then, in detail, in thermodynamic model using both canonical
and grand canonical ensembles.  
In particular we point out a different functional
relationship is more natural in certain physical situations.
Next we turn to cases where isoscaling is only approximate and how
such cases can be handled in the theoretical framework.  Summary and 
conclusions are presented in section X.

\section{Relating isoscaling parameter to chemical potential}
Equation (1) can be easily understood using a grand canonical model
for multifragmentation.  This allows us to relate $\alpha$ to chemical
potentials.

We assume that in a central collision, the two ions fuse, some pre-equilibrium
emission occurs and the fused system, because of two-body collisions
is heated up and begins to expand.  During the expansion composites are
formed.  As the expansion takes place interaction between compsites
rapidly fall off except for Coulomb interaction which can be taken care of
in an approximate way using the Wigner-Seitz apporoximation \cite{Bondorf1}.
In this expanded volume the break up of the dissociating system can
be calculated using laws of equilibrium statistical mechanics.  The 
calculation is particularly simple if a grand canonical ensemble
is used.

More about the grand canonical approximation will follow later but at this
stage let us quickly connect $\alpha$ to chemical potential
encountered in the grand canonical ensemble.  The cross-section
of the produced composite is given by $\sigma(N,Z)=C\langle n_{N,Z}\rangle$
where $C$ is a constant not provided by the model; $\langle n_{N,Z}\rangle$
is average multiplicity of the composite.  For system 1 characterized
by total charge $Z_0(1)$ and total mass $A_0(1)$ 
(total neutron number $N_0(1)=A_0(1)-
Z_0(1)$) and neutron and proton chemical potentials $\mu_n(1)$ and
$\mu_p(1)$ respectively this multiplicity is
\begin{eqnarray}
\langle n_{N,Z}(1)\rangle=e^{\beta\mu_n(1)N+\beta\mu_p(1)Z}\omega_{N,Z}
\end{eqnarray}
where $\beta$ is the inverse of temperature $T$ and
$\omega_{N,Z}$ is the one particle partition function of the
composite $N,Z$.

If in the second reaction the total charge is $Z_0(2)$, the total
mass is $A_0(2)$
but the conditions of the second reaction are similar to that of reaction 1 
and we expect the same temperature, then
\begin{eqnarray}
\frac{\sigma_2(N,Z)}{\sigma_1(N,Z)}\propto e^{\beta(\mu_n(2)-\mu_n(1))N+
\beta(\mu_p(2)-\mu_p(1))Z}
\end{eqnarray}
as the $\omega(N,Z)$ in the numerator cancels the $\omega(N,Z)$ in the
denominator.  Let $\delta\mu_n\equiv \mu_n(2)-\mu_n(1)$
and $\delta\mu_p\equiv \mu_p(2)-\mu_p(1)$.  A widely used relationship is
\begin{eqnarray}
\delta\mu_n\approx 4c_s
[(\frac{Z_0(1)}{A_0(1)})^2-(\frac{Z_0(2)}{A_0(2)})^2]
\end{eqnarray}
A corresponding relationship can be written down for $\delta\mu_p$.
It suffices to study any one and traditionally one examines $\delta\mu_n$.
In the following we will investigate $\delta\mu_n$ for various cases.

\section{$\delta\mu_n$ in a mean field model}
The concept of chemical potential is useful not only in problems
concerned with multifragmentation.  We first investigate the
chemical potential in mean field theory at finite temperature.
One might argue that this is a valid model at low temperatures
$T\leq 3$ MeV.  The caloric curve has been computed in this model
\cite {Sobotka1,Dasgupta1} and many interesting results were found.

For a nucleus with $N$ neutrons and $Z$ protons ($N+Z=A$) the symmetry
energy contributes to the binding energy a term:
$C_s\frac{(N-Z)^2}{(N+Z)}$.  To binding energy per particle it gives
$C_s\frac{(N-Z)^2}{A^2}$.  The term $C_s$ has its origin to both
kinetic and potential energy per particle so we separate $C_s$ into
two parts: $C_s=C_s(k.e)+C_s(p.e)$.

We consider asymmetric nuclear matter where proton charges are
switched off.
We can calculate both $C_s(k.e)$ and $C_s(p.e)$ in the Hartree-Fock
model interacting by Skyrme interaction.  For an asymmetric
nucleus $\rho_n$ and $\rho_p$ are different.  We have
\begin{eqnarray}
\rho=\rho_n+\rho_p \nonumber \\
\Delta=\frac{\rho_n-\rho_p}{\rho_n+\rho_p}=\frac{(N-Z)}{N+Z}=1-\frac{2Z}{A}
 \nonumber \\
\rho_n=\frac{\rho}{2}(1+\Delta) \nonumber \\
\rho_p=\frac{\rho}{2}(1-\Delta) \nonumber 
\end{eqnarray}

In the Hartree-Fock model at zero temperature, the kinetic energy per 
nucleon is given by
\begin{eqnarray}
\frac{K.E.}{A}=\frac{N}{A}\frac{3}{5}\frac{p_f^2(n)}{2m}+\frac{Z}{A}\frac{3}{5}
\frac{p_f^2(p)}{2m} \nonumber \\
=\frac{\rho_n}{\rho}\frac{3h^2}{10m}(\frac{3\rho_n}{8\pi})^{2/3}+
\frac{\rho_p}{\rho}\frac{3h^2}{10m}(\frac{3\rho_p}{8\pi})^{2/3}
\end{eqnarray}
Expanding the above in powers of $\Delta$ upto $\Delta^2$ we get
\begin{eqnarray}  
\frac{K.E.}{A}=\frac{3}{10m}h^2(\frac{3\rho}{16\pi})^{2/3}
(1+\frac{5}{9}\Delta^2)
\end{eqnarray}
This then identifies $C_s(k.e)$:
\begin{eqnarray}
C_s(k.e)=\frac{h^2}{6m}(\frac{3\rho}{16\pi})^{2/3}
\end{eqnarray}
Of course, in $\frac{K.E.}{A}$ terms of higher powers of $\Delta$
exist which will be small and are neglected.

For contribution $C_s(p.e)$ we start with the simplest potential 
energy density that will produce the correct saturation density, binding
energy, compressibility and symmetry energy coefficient \cite {Das1}:
\begin{eqnarray}
V(\rho_n,\rho_p)=\frac{A_u}{\rho_0}\rho_n\rho_p+\frac{A_l}{2\rho_0}
(\rho_n^2+\rho_p^2)+\frac{B}{\sigma+1}\frac{\rho^{\sigma+1}}{\rho_0^{\sigma}}
\end{eqnarray}
This will give for potential energy per particle:
\begin{eqnarray}
\frac{P.E.}{A}=A_u\frac{\rho_n\rho_p}{\rho\rho_0}+\frac{A_l}{2\rho\rho_0}
(\rho_n^2+\rho_p^2)+\frac{B}{\sigma+1}(\frac{\rho}{\rho_0})^{\sigma}
\end{eqnarray}
Writing $\rho_n, \rho_p$ in terms of $\rho$ and $\Delta$ we get
\begin{eqnarray}
\frac{P.E.}{A}=\frac{1}{4}\frac{\rho}{\rho_0}(A_u+A_l)+\frac{1}{4}
\frac{\rho}{\rho_0}(A_l-A_u)\Delta^2+
\frac{B}{\sigma+1}(\frac{\rho}{\rho_0})^{\sigma}
\end{eqnarray}
We identify $C_s(p.e)=\frac{1}{4}\frac{\rho}{\rho_0}(A_l-A_u)$.
With $A_u=-379.2$MeV,
$A_l=-334.4$MeV, $B$=303.9 MeV and $\sigma$=7/6, one gets for symmetric 
nuclear matter saturation density $\rho_0=0.16$fm$^{-3}$, B.E./A=16 MeV,
compressibility=210 MeV and at $\rho/\rho_0$=1, $C_s(p.e)$=11.2 MeV.
Together with $C_s(k.e)\approx 12.3$ Mev $C_s$ adds to
the total value of 23.5 MeV.

The Hartree-Fock energy of an orbital is given by
\begin{eqnarray}
\epsilon=p^2/2m+A_u(\rho_u/\rho_0)+A_l(\rho_l/\rho_0)+B(\rho/\rho_0)^{\sigma}
\end{eqnarray}
The value of $\mu_n$ is found by solving for a given $\rho_n$ and $\beta=1/T$
\begin{eqnarray}
\rho_n=\frac{8\pi}{h^3}\int_0^{\infty}\frac{p^2dp}{\exp[\beta(\epsilon_n-
\mu_n)]+1}
\end{eqnarray}
In the model pursued here, the potential part in $\epsilon$
is constant for given densities and is given by $K=\frac{\rho}{2\rho_0}
(A_u+A_l)+\frac{\rho(A_l-A_u)}{2\rho_0}\Delta+
B(\frac{\rho}{\rho_0})^{\sigma}=\frac{\rho}{2\rho_0}(A_u+A_l)+2C_s(p.e)\Delta+
B(\frac{\rho}{\rho_0})^{\sigma}$.
For zero temperature, the chemical potential (at zero temperature 
$\mu=\epsilon_f$) is $\mu=\frac{p_f^2}{2m}+K$.  We first consider
zero temperature.  This will be followed by the finite temperature case.

The change in neutron chemical potentials for two nuclei: one with
$\Delta_2=1-\frac{2Z_2}{A_2}$ and another with $\Delta_1=1-\frac{2Z_1}{A_1}$
is
\begin{eqnarray}
\delta\mu_n=\mu_n(2)-\mu_n(1)
=\frac{p_f^2(n)}{2m}(2)-\frac{p_f^2(n)}{2m}(1)+4C_s(p.e)
(\frac{Z_1}{A_1}-\frac{Z_2}{A_2})
\end{eqnarray}
The kinetic term is:
\begin{eqnarray}
\frac{p_f^2(n)}{2m}(2)-\frac{p_f^2(n)}{2m}(1)=\frac{h^2}{2m}
(\frac{3}{8\pi})^{2/3}[(\frac{\rho}{2})^{2/3}(1+\Delta_2)^{2/3}-(\frac{
\rho}{2})^{2/3})^(1+\Delta_1)^{2/3}]
\end{eqnarray}
Expanding the above to the lowest order in $\Delta$ 
\begin{eqnarray}
\frac{p_f^2(n)}{2m}(2)-\frac{p_f^2(n)}{2m}(1)
\approx \frac{h^2}{6m}(\frac{3\rho}{
16\pi})^{2/3}4(\frac{Z_1}{A_1}-\frac{Z_2}{A_2})
\end{eqnarray}
Together then we get 
\begin{eqnarray}
\delta\mu_n\approx 4.0*C_s(\frac{Z_1}{A_1}-\frac{Z_2}{A_2})
\end{eqnarray}
the approximate nature of the above equation arises because contributions
from kinetic energy have been retained to lowest orders in $\Delta$.
Note the difference of the above equation from the generally used relation of 
eq.(4).

Consider now finite temperature mean field theory.  The contribution to
$\delta\mu_n$ from potential energy does not change.  But the contribution
to $\delta\mu_n$ from kinetic energy will change. 
An approximate answer, quite accurate upto 6 MeV
temperature, is $[\frac{p_f^2(n)}{2m}(2)-\frac{p_f^2(n)}{2m}(1)](1+
\frac{\pi^2T^2}{12e_0(1)e_0(2)})$ with $e_0(1)=\frac{p_f^2(n)}{2m}(1)$
and $e_0(2)=\frac{p_f^2(n)}{2m}(2)$.  But it is easy to get
an accurate answer for all temperatures numerically.
Fig.1 shows that in this model eq.(16) works better than eq.(4).

One can do refinements to this model.  For example as the temperature
increases, the nucleus will expand \cite{Sobotka1,Dasgupta1} 
which will cause some quantitative changes.  But we will not
pursue these finer details.  

\section{$\delta\mu_n$ in thermodynamic multifragmentation 
models: canonical and grandcanonical}
We now go back to the multifragmentation model that we briefly alluded to
in section II.  Assume that the system with $A_0$ nucleons and $Z_0$
protons has temperature $T$, has expanded to a higher than normal volume 
and the partitioning into different composites can be calculated according
to equilibrium statistical mechanics.  In a canonical model, the partitioning
is done such that all partitions have the correct $A_0, Z_0$ (equivalently
$N_0, Z_0$).  Details of the implementation of the canonical model
can be found elsewhere \cite{Das2}; here we give the essentials
necessary to follow the present work.

The canonical partition function is given by
\begin{eqnarray}
Q_{N_0,Z_0}=\sum\prod \frac{\omega_{I,J}^{n_{I,J}}}{n_{I,J}!}
\end{eqnarray}
Here the sum is over all possible channels of break-up (the number of such
channels is enormous) which satisfy $N_0=\sum I\times n_{I,J}$
and $Z_0=\sum J\times n_{I,J}$; $\omega_{I,J}$ 
is the partition function of one composite with
neutron number $I$ and proton number $J$ respectively and $n_{I,J}$ is
the number of this composite in the given channel.  
The one-body partition
function $\omega_{I,J}$ is a product of two parts: one arising from
the translational motion of the composite and another from the
intrinsic partition function of the composite:
\begin{eqnarray}
\omega_{I,J}=\frac{V_f}{h^3}(2\pi mT)^{3/2}A^{3/2}\times z_{I,J}(int)
\end{eqnarray}
Here $A=I+J$ is the mass number of the composite and
$V_f$ is the volume available for translational motion; $V_f$ will
be less than $V$, the volume to which the system has expanded at
break up. We use $V_f = V - V_0$ , where $V_0$ is the normal volume of  
nucleus with $Z_0$ protons and $N_0$ neutrons.  In this calculation we
have used a fairly typical value $V=6V_0$.

The probability of a given channel $P(\vec n_{I,J})\equiv P(n_{0,1},
n_{1,0},n_{1,1}......n_{I,J}.......)$ is given by
\begin{eqnarray}
P(\vec n_{I,J})=\frac{1}{Q_{N_0,Z_0}}\prod\frac{\omega_{I,J}^{n_{I,J}}}
{n_{I,J}!}
\end{eqnarray}
The average number of composites with $I$ neutrons and $J$ protons is
seen easily from the above equation to be
\begin{eqnarray}
\langle n_{I,J}\rangle=\omega_{I,J}\frac{Q_{N_0-I,Z_0-J}}{Q_{N_0,Z_0}}
\end{eqnarray}
The constraints $N_0=\sum I\times n_{I,J}$ and $Z_0=\sum J\times n_{I,J}$
can be used to obtain different looking but equivalent recursion relations
for partition functions.  For example
\begin{eqnarray}
Q_{N_0,Z_0}=\frac{1}{N_0}\sum_{I,J}I\omega_{I,J}Q_{N_0-I,Z_0-J}
\end{eqnarray}
These recursion relations allow one to calculate $Q_{N_0,Z_0}$ 

We list now the properties of the composites used in this work.  The
proton and the neutron are fundamental building blocks 
thus $z_{1,0}(int)=z_{0,1}(int)=2$ 
where 2 takes care of the spin degeneracy.  For
deuteron, triton, $^3$He and $^4$He we use $z_{I,J}(int)=(2s_{I,J}+1)\exp(-
\beta E_{I,J}(gr))$ where $\beta=1/T, E_{I,J}(gr)$ is the ground state energy
of the composite and $(2s_{I,J}+1)$ is the experimental spin degeneracy
of the ground state.  Excited states for these very low mass
nuclei are not included.  
For mass number $A=5$ and greater we use
the liquid-drop formula.  For nuclei in isolation, this reads ($A=I+J$)
\begin{eqnarray}
z_{I,J}(int)=\exp\frac{1}{T}[W_0A-\sigma(T)A^{2/3}-\kappa\frac{J^2}{A^{1/3}}
-C_s\frac{(I-J)^2}{A}+\frac{T^2A}{\epsilon_0}]
\end{eqnarray}
The derivation of this equation is given in several places
\cite{Bondorf1,Das2}
so we will not repeat the arguments here.  The expression includes the 
volume energy, the temperature dependent surface energy, the Coulomb
energy and the symmetry energy.  The term $\frac{T^2A}{\epsilon_0}$
represents contribution from excited states
since the composites are at a non-zero temperature.  

For most of the calculations here
the dissociating system is $N_0=93, Z_0=75, A_0=168$ for reaction 1.
For reaction 2, the dissociating system is $N_0=111, Z_0=75, A_0=186$.
These will represent $^{112}$Sn+$^{112}$Sn central collisions
and $^{124}$Sn+$^{124}$Sn central collisions after pre-equilibrium
particles are emitted.  These two systems have received much attention
in the past.
We also have to state which nuclei are included in computing $Q_{N_0,Z_0}$ 
(eq.(17)).
For $I,J$, (the neutron and the proton number)
we include a ridge along the line of stability.  The liquid-drop
formula above also gives neutron and proton drip lines and 
the results shown here include all nuclei within the boundaries.

The long range Coulomb interaction between
different composites can be included in an approximation called
the Wigner-Seitz approximation.  We incorporate this following the
scheme set up in \cite{Bondorf1}.  

Computations of observables with the canonical model can be done without
an explicit use of a chemical potential.  We can, however, compute
the chemical potential using the thermodynamic identity
$\mu=(\partial F/\partial n)_{V,T}$ \cite{Reif}.  We know the values
of $Q_{N_0,Z_0}, Q_{N_0-1,Z_0}$ and $Q_{N_0,Z_0-1}$.  Since free energy $F$
is just $-T$ln$Q$ we compute $\mu_n$ from 
$-T($ln$Q_{N_0,Z_0}$-ln$Q_{N_0-1,Z_0}$) and $\mu_p$ from 
$-T($ln$Q_{N_0,Z_0}$-ln$Q_{N_0,Z_0-1}$).

We now briefly review the grand canonical model.  For finite systems such 
as considered here it is inferior to the canonical model but is easier to
implement.  If the numbers of neutrons and protons in the dissociating
system are $N_0$ and $Z_0$ respectively, the ensemble contains not
only these but many others but the average value can be constrained to
be $N_0$ and $Z_0$.  The chemical potentials $\mu_n$ and $\mu_p$ serve
to fix the average numbers.  If the neutron chemical potential is $\mu_n$
and the proton chemical potential is $\mu_p$, then statistical equilibrium
implies that the chemical potential of a composite
with $N$ neutrons and $Z$ protons is $\mu_nN+\mu_pZ$.  The following are the
relevant equations for us.  The average number of composites with $N$ neutrons
and $Z$ is
\begin{eqnarray}
\langle n_{N,Z}\rangle=e^{\beta\mu_nN+\beta\mu_pZ}\omega_{N,Z}
\end{eqnarray}
There are two equations which determine $\mu_n$ and $\mu_p$.
\begin{eqnarray}
N_0=\sum Ne^{\beta\mu_nN+\beta\mu_pZ}\omega_{N,Z}
\end{eqnarray}
\begin{eqnarray}
Z_0=\sum Ze^{\beta\mu_nN+\beta\mu_pZ}\omega_{N,Z}
\end{eqnarray}
The sum here is over all nuclei within drip lines whose $(N,Z)$ do not exceed
$(N_0,Z_0)$ since there can not be a composite whose $N,Z$ exceed those
of the system from which it emerges.

We want to point out the following feature of the grand canonical model.  
In all $\omega_{N,Z}$'s in the sum in the above two equations,
there is one common value  for $V_f$ (see eq.(18)).  We really solve 
for $N_0/V_f$
and $Z_0/V_f$.  The values of $\mu_n$ or $\mu_p$ will not change if we, say,
double $N_0, Z_0$ and $V_f$ simultaneously provided the number of
terms in the sum is unaltered.  We then might as well say that when
we are solving the grand canonical equation we are really solving for
an infinite system (because we know that fluctuations will become
unimportant) but this infinite system can break up into only certain
kinds of species as are included in the above two equations.  Which
composites are included in the sum is an important physical ingredient
in the model but intensive quantities like $\beta,\mu$ depend not
on $N_0, Z_0$ but on $N_0/V_f$ and $Z_0/V_f$.  To apply the grand canonical
model to finite systems after solving for $\mu$'s we plug in the value
of $V_f$ that would be appropriate for the system $N_0,Z_0$.
If the system which
we are investigating is small, experimental data may show
substantial deviations from the grand canonical
model as we will verify later.

For later application, we will also use
a slightly different version of the above
equations \cite{Botvina1}.  We label two other chemical potentials:
$\mu$ (fixes baryon number) and $\nu$ (fixes total charge):
\begin{eqnarray}
A_0=\sum Ae^{\beta\mu A+\beta\nu Z}\tilde{\omega}_{A,Z}
\end{eqnarray}
\begin{eqnarray}
Z_0=\sum Ze^{\beta\mu A+\beta\nu Z}\tilde{\omega}_{A,Z}
\end{eqnarray}
Here $\mu=\mu_n$, $\nu=\mu_p-\mu$ and $\tilde{\omega}_{A,Z}=\omega_{N,Z}$ with
$A=N+Z$.

For $A_0=168, Z_0=75$ and $A_0=186, Z_0=75$ the $\mu_n$'s of the
grand canonical model are compared with those derived from the
canonical model in Fig.2.  The values are very close.  
We also compare the $\delta\mu_n$ in
the two models.

\section{Computation of $\delta\mu_n$ as $T\rightarrow 0$}
The computation of $\mu$ whether in canonical or grand canonical 
requires solving complicated equations.  However it may simplify as 
$T\rightarrow 0$ (see also \cite{Botvina1}).  We try this in the
canonical model first.

As $T\rightarrow 0$ the translational degree of freedom can be considered
frozen.
Let $A_0,Z_0$ be stable against spontaneous dissociation (if $A_0,Z_0$
is one of the nuclei within neutron and proton drip lines then it can
not spontaneusly decay into a neutron(proton) plus a daughter;
usually the only other channel one needs to check is an alpha plus
daughter).  As $T\rightarrow 0$, the system will drop to the ground state
of $A_0,Z_0$ and
we will have (eq.(17)) $Q_{N_0,Z_0}\rightarrow
\omega_{N_0,Z_0}$ with $\omega_{N_0,Z_0}$ now given by
$\exp(-\frac{1}{T}E_{gr}(N_0,Z_0))$.  Consequently we get $\mu_n(N_0,Z_0)=
E_{gr}(N_0,Z_0)-E_{gr}(N_0-1,Z_0)$.  This result is of course
physically meaningful: $-\mu_n$ is simply the separation energy required
to free a neutron from bound nucleus $N_0,Z_0$.  Thus
\begin{eqnarray}
\mu_n(N_0,Z_0) & = & -W_0+term1+term2  \\
term1 & = & \sigma[A_0^{2/3}-(A_0-1)^{2/3}]+\kappa Z_0^2[A_0^{-1/3}-
(A_0-1)^{-1/3}]  \\
term2 & = & C_s[\frac{(N_0-Z_0)^2}{A_0}-\frac{(N_0-1-Z_0)^2}{A_0-1}]
\end{eqnarray}
For us $term2$ is of greater significance.  It can be rewritten as
\begin{eqnarray}
term2 & = & C_s(1-\frac{4Z_0^2}{A_0(A_0-1)}) \nonumber \\
      & \approx & C_s(1-\frac{4Z_0^2}{A_0^2})
\end{eqnarray}
For $\delta\mu_n$ we need to take the difference between two
chemical potentials.  $Term1$ of eq.(28) contributes little in the difference
and thus we end up with familiar eq.(4), 
i.e., $\delta\mu_n=4*[(\frac{Z_1}{A_1})^2-(\frac{Z_2}{A_2})^2]$.

Eq.(16) can also be obtained but as an approximation to eq.(30).  We can
rewrite eq.(30) as
\begin{eqnarray}
term2=2C_s-C_s\frac{1+4Z_0}{A_0}-C_s\frac{(N_0-Z_0-1)^2}{A_0(A_0-1)}
\end{eqnarray}
For the examples we are using, $A_0=168(186)$ and $Z_0=75$, the third
term in the right hand side of the above equation is much less 
important than the second term.
If we neglect the third term and as before also term1
 we end up with eq.(16), i.e., $\delta\mu_n=4*(Z_1/A_1-Z_2/A_2)$.

Let us see if we can get a sensible answer in the grand canonical model.
We will use the the alternative forms eqs (26) and (27): $\mu$ controls
the baryon number and $\nu$ the total charge.
So long as we maintain the general form of eqs.(26) and (27), that is,
include in the sum all particle stable nuclei with $A\leq A_0, Z\leq Z_0$
the limits of $\mu$ and $\nu$ are very difficult to obtain even at the zero
temperature limit.  In particular we can not have mutiplicity 1
in the ground state of the dissociating system $A_0,Z_0$ and
zero occupation in all other composites in the sum in eqs. (26) and
(27) (see, however, \cite{Botvina1}).  
Exclusive occupation in the ground state of $A_0,Z_0$ can
be achieved in the canonical model but not in the
grand canonical ensemble as this violates a fluctuation equation.
We will deal with that equation but let us look at this first in
a more pedestrian fashion.  At zero temperature the translational degree
can be considered frozen.  From eqs. (26) and (27), to get mutiplicity
1 we need $(\mu A_0+\nu Z_0-E_{gr}(A_0,Z_0)) \rightarrow 0$ as
$\beta\rightarrow \infty$.  This alone is not enough to determine $\mu$ or
$\nu$ but we also require $\mu A_0+\nu Z-E_{gr}(A_0,Z)$ to
maximise at $Z=Z_0$ so that at other values of $Z$ the difference is
negative and occupation in $Z'$s other than $Z_0$ will go to 0 as 
$\beta\rightarrow \infty$.  The maximisation condition gives
\cite{Botvina1}
\begin{eqnarray}
\nu=-4C_s\frac{A_0-2Z_0}{A_0}+\kappa\frac{2Z_0}{A_0^{1/3}}
\end{eqnarray}
Having determined $\nu$, the value of $\mu$ can be found from
$\mu A_0+\nu Z_0-E_{gr}(A_0,Z_0)=0$.  This precudure ensures that
$\mu A_0+\nu Z-E_{gr}(A_0,Z)$ is negative for $Z<Z_0$ and hence
as $\beta\rightarrow \infty$ the occupation in composites labelled
by $A_0,Z$ with $Z<Z_0$ will go to zero.  But this does not guarantee
that $\mu A+\nu Z-E_{gr}(A,Z)$ will be less than zero for all $A$'s
less than $A_0$ with $Z$'s less than $Z_0$ that are in the sum
of eqs. (26) and (27).  In fact they are not all negative and whenever 
they are postive, multiplicities for those $(A,Z)$'s blow up.  Another
way of understanding this is to realise that for sole occupation
in $(A_0,Z_0)$ there are three conditions to be met: $\mu A+\nu Z
-E_{gr}(A,Z)$ must (1) go to 0 at $A_0, Z_0$, (2) must maximise
as a function of $Z$ at $A_0,Z_0$ and (3) must maximise as a function
of $A$ at $A_0,Z_0$.  With only two parameters $\mu$ and $\nu$ this
can not be achieved.

We can also deduce the impossibility of exclusive occupation in the 
ground state of $A_0,Z_0$ from very general arguments
about fluctuations.  It is easy to
prove this when there is only one kind of particle (eqs.(19) to (21)
in \cite{Das2}).  With 2 kinds of particles, neutrons and protons and
hence 2 chemical potentials $\mu$ and $\nu$ the notation gets complicated. 
Quite generally, the equation for 
the grand canonical partition function, when there are
many species $i$ which are non-interacting, is given by
\begin{eqnarray}
Z_{gr.can}=\prod_i(\sum_{n_i=0}^{\infty}(e^{\beta\mu_i})^{n_i}z_{n_i}(i))
\end{eqnarray} 
In our case $i$ stands for both $a$ and $z$, the composite mass and charge;
$z_{n_i}(i)$ is the canonical partition function of $n_i$ particles of
type $i$; $n_i$ goes from 0 to $\infty$ as we are constructing a grand 
partition function.  We have $\mu_i=\mu a+\nu z$.  We need not specify the
functional form of $z_{n_i}(i)$.

Eq.(34) can be rewritten as
\begin{eqnarray}
Z_{gr.can}=\sum_{m=0}^{\infty}e^{\beta\mu m}\tilde{z}_m
\end{eqnarray}
where we have absorbed the factors $e^{\beta\nu z}$
in $\tilde{z}_m$ which is
quite complicated but it only contains partition functions with total
particle number $m$.  We clearly have (fluctuation equation):
\begin{eqnarray}
\langle m^2\rangle-\langle m\rangle^2
=\frac{1}{\beta^2}\frac{\partial^2}{\partial^2\mu}\ln Z_{gr}.
\end{eqnarray}
In our case $\ln Z_{gr}$ is particularly simple:
\begin{eqnarray}
\ln Z_{gr}=\sum e^{\beta\mu A+\beta\nu Z}\tilde{\omega}_{A,Z}
\end{eqnarray}
and
\begin{eqnarray}
\frac{1}{\beta^2}\frac{\partial^2}{\partial^2\mu}\ln Z_{gr}=
\sum_{A=1,A_0}A^2\langle n_A\rangle
\end{eqnarray}
If $\langle n_A\rangle=1$ for $A=A_0$ and 0 for all others then
$\langle m^2\rangle -\langle m\rangle^2=A_0^2$ but this contradicts
the assumption that the only occupation is for $m=A_0$ in which
case the fluctuations would have been 0.

In spite of this conceptual difficulty, the expression for $\mu_n$
derived from the grand canonical ensemble in \cite{Botvina1} and
that derived here from canonical ensemble are not that different.
In particular $\delta\mu_n$ will be practically the same.  For
completeness we write the two, one after the other.  That in 
\cite{Botvina1} is
\begin{eqnarray}
\mu_n=-W_0+\frac{\sigma}{A_0^{1/3}}-\kappa\frac{Z_0^2}{A_0^{4/3}}
+C_s[1-(\frac{2Z_0}{A_0})^2]
\end{eqnarray}
whereas we get
\begin{eqnarray} 
\mu_n=-W_0+\sigma[A_0^{2/3}-(A_0-1)^{2/3}]+\kappa Z_0^2[A_0^{-1/3}-
(A_0-1)^{-1/3}]+C_s[1-(\frac{2Z_0}{A_0})^2]
\end{eqnarray}  

\section{$\delta\mu_n$ in more general case}
The contribution of the symmetry energy 
to $\delta\mu_n$ in the more general case 
will be function of temperature and not a constant as implied in 
eq.(4).  This can most readily be seen by analytically deriving
$\delta\mu_n$ at high temperature.  Let us derive this in the 
canonical ensemble first.  At very high temperature we will get only
neutrons and protons and so eq.(17) becomes particularly simple:
\begin{eqnarray}
Q_{N_0,Z_0}\rightarrow \frac{\omega_{1,0}^{N_0}}{N_0!}
\frac{\omega_{0,1}^{Z_0}}{Z_0!}
\end{eqnarray}
The formula $\mu_n=-T[\ln Q_{N_0,Z_0}-\ln Q_{N_0-1,Z_0}]$ leads to
$\mu_n(N_0,Z_0)=-T\ln\frac{\omega_{1,0}}{N_0}$ and hence
$\delta\mu_n=Tln(N_2/N_1)$ and thus not a function of the 
symmetry energy at all.  It is easy to verify that in this 
high temperature limit
the grand canonical ensemble gives identical answers. 
From $T=0$ towards a large value of $T$
this must happen gradually and so $\delta\mu_n$ must be an evolving
function of $T$.  

We are unable to derive a simple formula for $\delta\mu_n$ for a 
general $T$.  The reasons for this failure are obvious enough.
Eqs.(24) and (25) are highly non-linear and thus the grand canonical
ensemble provides no simple leads.  The relevant equations for
the canonical model are easy to compute on a computer but otherwise
are quite non-transparent.  In Fig.3 we show 
as a function of temperature (relates to beam energy of collision)
numerical results for
$\delta\mu_n$ for the two systems studied here.  Two results are
shown; one with $C_s=23.5$ MeV and another with $C_s=15$ MeV.  
Extracting the value of $C_s$ from $\delta\mu_n$ using
eq.(4) is difficult because of temperature and hence
beam energy dependence.  Notice that at high temperature,
the curves obtained with different values of $C_s$ approach one another.
They will converge to one line which will rise linearly with $T$.

A very interesting aspect of the experimental data is the 
behavior of $\delta\mu_n$ when $Z_2/A_2$
varies for fixed $Z_1/A_1$ or vice versa.  In the thermodynamic model
this depends on the beam energy.  There is no such dependence
in eq.(4).  The simple prediction of eq.(4) is compared with
thermodynamic model predictions in Fig.4.  In the model, for fixed
$Z_1/A_1$, $\delta\mu_n$ is still approximately linear with $(Z_2/A_2)^2$
although deviation can be seen by eye-estimation.  It is also noticed
that use of eq.(4) to estimate $C_s$ from $\delta\mu_n$ 
will underestimate the value of $C_s$ at low temperature (4 or 5 MeV)
but will overestimate the value at high temperature, for example, at 7 MeV.
 
Experiments directly measure $\alpha=\beta\mu_n$ rather than $\mu_n$
and we plot $\alpha$ from the thermodynamic model as a function of 
temperature in Fig.5.  We also
obtain $\alpha$ from eq.(4) and compare.  Eq.(4) gives a $1/T$ dependence
but the fall with $T$ is much slower in the thermodynamic model
at higher temperature.  In the thermodynamic model $\alpha$ would
reach asymptotically a constant value $\ln \frac{N_2}{N_1}$.  This
difference in behavior between the two predictions can 
be ascertained in experiments.

There has been much activity in recent times relating $\alpha$ to
$C_s$ \cite{Shetty1,Shetty2,Souliotis}.  An approximate derivation of
eq.(4) from the expanding excited source (EES) model can be found
in \cite{Tsang2}.  Attempts to obtain this from antisymmetrised
molecular dynamics can be found in \cite{Ono}.  Temperature dependence
of symmetry energy was discussed by Li and Chen \cite{Li}.

\section{Deviation from Isoscaling: explanation from canonical model }
We now discuss isoscaling and possible deviations from it.
It will become clear in this and the following section that,
from the point of view of theory,
isoscaling can work very well when $N/N_0$ and $Z/Z_0$ are small
($\le 0.35)$.  Many experimental data fall in this range 
\cite{Xu,Tsang1,Tsang2} and isoscaling is one robust feature 
to emerge in recent experimental intermediaite energy
heavy ion collisions.  If we now extend these observations to
larger composites deviations are to be expected.
Will the grand canonical and canonical results for $R_{21}$ always agree?
The answer is no; it depends upon the size of the dissociating system
($N_0,Z_0$), more precisely, upon the fractions $N/N_0$ and $Z/Z_0$.  The
grand canonical model always predicts isoscaling.  In this model
\begin{eqnarray}
R_{21}&=&C\frac{\langle n_{N,Z}(2)\rangle}{\langle n_{N,Z}(1)\rangle} 
\nonumber \\
      &=&C\exp (\alpha N+\beta Z)
\end{eqnarray}
where we have used eq.(2) and the advantage that with similar beam conditions
the factors $\omega_{N,Z}$ in the denominator and the numerator cancel
each other out.  Therefore, in the grand canonical approximation, the
slopes of $\ln R_{21}$ as a function of $N$ for fixed Z (a) will never deviate
from a straight line and (b) for different fixed $Z$'s the slopes will not
change.  In many experiments where the sizes of the composites \cite{Mocko1} 
encompass from small to large 
this is not true.  In Fig.6
we compare some experimental data with a grand canonical calculation.
Experimental details can be found in \cite{Mocko1,Mocko2,Mocko3}.
In the experiment, reaction 1 is $^{58}$Ni on $^9$Be and reaction 2
is $^{64}$Ni on $^9$Be.  For the grand canonical calculation, for
reaction 1 the dissociating system is taken to be $^{58}$Ni+ $^9$Be
($N_0=35, Z_0=32$) and for reaction 2 the dissociating system is taken
to be $^{64}$Ni+ $^9$Be ($N_0=41, Z_0=32$).  All composites between
drip lines are included as detailed in section IV with the highest
values of $N,Z$ terminating at $N_0,Z_0$.  The limitations of
the grand canonical model are very obvious in the figure.  The slopes
of $\ln R_{21}$ differ from straight lines for large $Z$ and as well
the same slope for all $Z$ does not fit.  In these calculations the
same temperature was assumed for all composites.  The actual situation
may be more complicated requiring a different temperature for higher
composites.  However the deviation from linearity will require even
further complications if one insists on using the grand canonical model
to fit the data.  From the point of view of theory, however, for the
emitting systems in these cases, the use of the grand canonical
approximation for the emissions of heavier composites is not valid.

The canonical model does not impose these restrictions.  Now (see eq.(20))
\begin{eqnarray}
R_{21}=C\frac {Q_{N_0-N,Z_0-Z}(2)}{Q_{N_0,Z_0}(2)}/
\frac {Q_{N_0-N,Z_0-Z}(1)}{Q_{N_0,Z_0}(1)}
\end{eqnarray}
This formula is not transparent at all but produces deviations
from isoscaling for small systems.

Let us show the results of the canonical model calculation
for the same case as above: Ni on Be.
The parameters for the
calculations are the same as used for the grand 
canonocal model: the same temperature, the same freeze-out volume and 
the same composites included in building the respective partition
functions. But the canonical calculations (Fig.7) are significantly
different from the grand canonical results (Fig.6) and much closer to
experimental data. 

\section{Further observations about differences between the canonical
and grand canonical Results}

Instead dealing directly with 
$R_{21}\propto \frac{\langle n_{N,Z}(2)\rangle}
{\langle n_{N,Z}(1)\rangle}$ let us investigate a simpler task:
that of relating $\langle n_{N,Z}\rangle$ in canonical with
that in grand canonical.  The formula for $\langle n_{N,Z}\rangle$, 
eq.(20), respects constancy of total neutron and proton numbers
in the partition function
and in this regard is superior to eq.(23).  If $N/N_0$ or/and
$Z/Z_0$ are not small (for some cases in Figs.6 and 7
they are $\ge 0.5$), eq.(20)
can give quite different results from those given by eq.(23).
This is so even when the chemical potentials given by the two models
are very close.  If the chemical potentials in the two ensembles are
nearly the same (as found in all our examples so far) that merely
guarantees that
\begin{eqnarray}
\frac{Q_{N_0-1,Z_0}}{Q_{N_0,Z_0}}\approx e^{\beta\mu_n} \nonumber \\
\frac{Q_{N_0,Z_0-1}}{Q_{N_0,Z_0}}\approx e^{\beta\mu_p}
\end{eqnarray}
Here the left hand side is given by the canonical model (and leads
to the definition of fugacity and 
chemical potential in the canonical model) and
the right hand side is computed from the grand canonical model
(eqs.(24) and (25)).  However, for $\langle n_{N,Z}\rangle$ to be nearly
the same in the two ensembles we require
\begin{eqnarray}
\frac{Q_{N_0-N,Z_0-Z}}{Q_{N_0,Z_0}}\approx e^{\beta\mu_n N+\beta\mu_p Z}
\end{eqnarray}
We show below how the left hand side of the above equation can lead to
the right hand side when $N/N_0$ and $Z/Z_0$ are small but are expected
to deviate when they are not small.  To proceed let us call
$\frac{Q_{A-1,B}}{Q_{A,B}}$=neutron fugacity of the system $A,B$ and
$\frac{Q_{A,B-1}}{Q_{A,B}}$=proton fugacity of the system $A,B$.
Rewrite the left hand side of eq.(45) as
\begin{eqnarray}
\frac{Q_{N_0-N,Z_0-Z}}{Q_{N_0,Z_0}}=\frac{Q_{N_0-N,Z_0-Z}}{Q_{N_0,Z_0-Z}}
\times\frac{Q_{N_0,Z_0-Z}}{Q_{N_0,Z_0}}
\end{eqnarray}
The first term on the right hand side can be rewritten as a product
of $N$ terms involving neutron fugacities and the second term as a 
product of $Z$ terms involving proton fugacities.
\begin{eqnarray}
\frac{Q_{N_0-N,Z_0-Z}}{Q_{N_0,Z_0-Z}}=\frac{Q_{N_0-N,Z_0-Z}}{Q_{N_0-N+1,Z_0-Z}}
\frac{Q_{N_0-N+1,Z_0-Z}}{Q_{N_0-N+2,Z_0-Z}}.....\frac{Q_{N_0-1,Z_0-Z}}
{Q_{N_0,Z_0-Z}}
\end{eqnarray}
The first term in the right hand side of the above equation is the 
neutron fugacity of the system $N_0-N+1,Z_0-Z$, the second term is
the neutron fugacity of the system $N_0-N+2,Z_0-Z$ and so on; finally ending
with neutron fugacity of the system $N_0,Z_0-Z$.  If $N$ is negligibly
small compared to $N_0$ and also $Z$ is negligibly small compared to
$Z_0$ then each of these terms can be approximated by neutron fugacity
of the system $N_0,Z_0$ (for $N\approx$ 20 and $Z\approx$ 20 in Fig.7,
this leads to gross errors) leading to 
$(\frac{Q_{N_0-1,Z_0}}{Q_{N_0,Z_0}})^N$. 
Equating $\frac{Q_{N_0-1,Z_0}}{Q_{N_0,Z_0}}$ to $e^{\beta\mu_n}$
we get the factor $e^{\beta\mu_n N}$ of eq. (45). (Even this can
introduce significant error for $N\approx$20.)  It is clear also how
$e^{\beta\mu_pZ}$ can arise by resolving $\frac{Q_{N_0,Z_0-Z}}{Q_{N_0,Z_0}}$
into $Z$ proton fugacities.

The connection between the two sides in eq. (45) can also be
established using a saddle-point approximation \cite{Das3} but we will
not pursue this any further.

It then follows that although at low $N,Z$ canonical and grand canonical
calculations can agree they will diverge when $N,Z$ grow.  This is
highlighted in Fig.8.

\section{Discussion about secondary decays}
Before comparing with data (as is done in Fig.7) one needs to investigate
the effects of sequential decay on calculated $R_{21}$.  The
multiplicities $\langle n_{I,J}\rangle$ refer to populations of nuclei
at finite temperatures.  Nuclei at finite temperatures can decay by 
particle emissions and values of $\langle n_{I,J}\rangle$
will change.  However, because one is comparing ratios, the effect on
$R_{21}$ may be less drastic.  We have included the contributions
from excited states through a factor $T^2A/\epsilon_0$
in eq.(22).  This overestimates contributions to the partition
function from excited states and a cut-off will be necessary \cite{Koonin}.
This requires detailed work which we have not carried out.
In investigating isoscaling using antisymmetrised molecular dynamics,
Ono et al. \cite{Ono2} 
find that the effect of secondary decay is to decrease $\alpha$
to $\alpha/2$.  This means the experimental value of $logR_{21}$ should
be compared to $log\sqrt{n(2)/n(1)}$ rather than to $log[n(2)/n(1)]$.  We have
applied this ``empirical'' correction to the Ni on Be case in Fig.9.
We now use temperature $T$=5 MeV. This leads to a steeper rise 
(compared to the $T$=8 MeV).  But after the correction the rise decreases
to a value more compatible with experiments.  It now corresponds roughly to 
``uncorrected'' $T$=8 MeV calculation.

\section{Summary}
This work has addressed two issues.  How well does eq.(4), a much
used formula, fit results from calculations obtained from
well-known models of nuclear dissociation?  We find the fit approximately
correct though not exact (the mean field model gives a different
formula but the mean field model is not a model for multifragmentation).
A more important question is: how do we handle cases when deviations
from isoscaling are significant.  Preliminary results reported
here are very encouraging.  Statistical multifragmentation can
explain the results but one must give up using the grand canonical
ensemble and instead use models that strictly conserve neutron and proton
numbers.   

\section{Acknowledgement}
This work is supported by Natural Sciences and Engineering Research
Council Of Canada and National Science Foundation under Grant No. PHY-0606007.

\newpage

\begin{figure}
\includegraphics[width=5.5in,height=4.5in,clip]{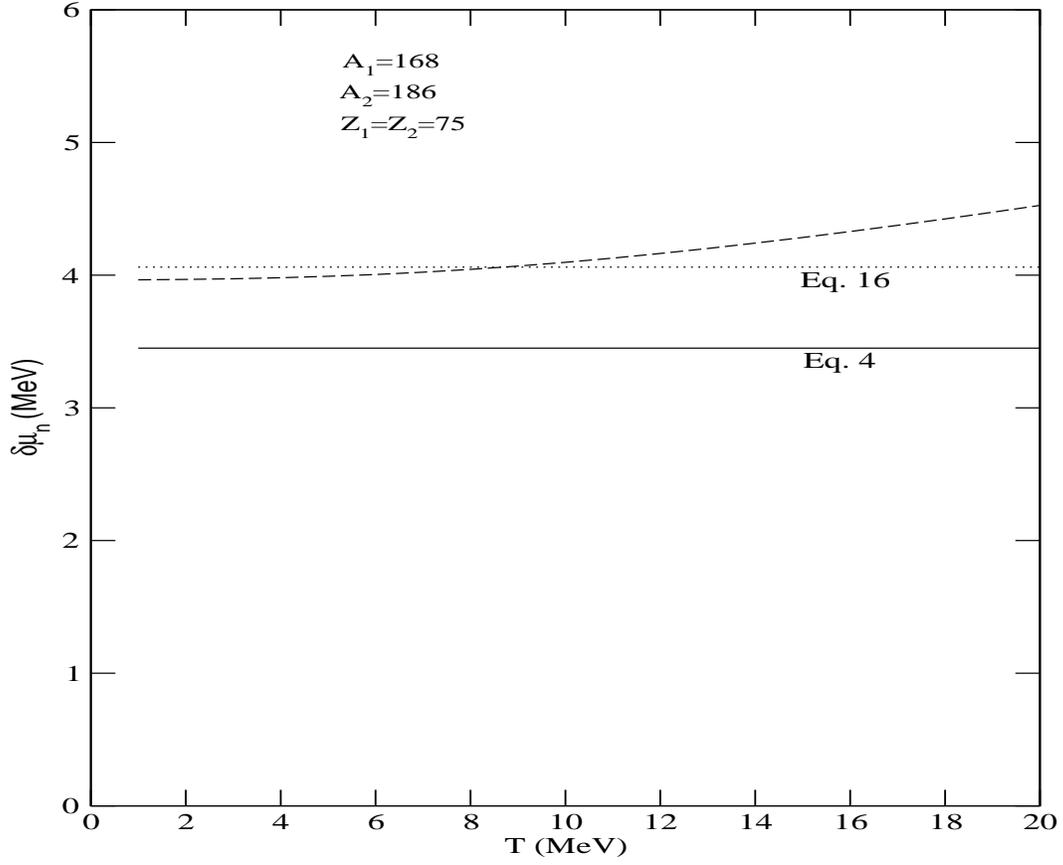}
\caption{ Mean-field model calculation (dashed curve) for $\delta\mu_n$
compared with simple versions (eq.(16) and eq.(4)).}
\end{figure}

\begin{figure}
\includegraphics[width=5.5in,height=4.5in,clip]{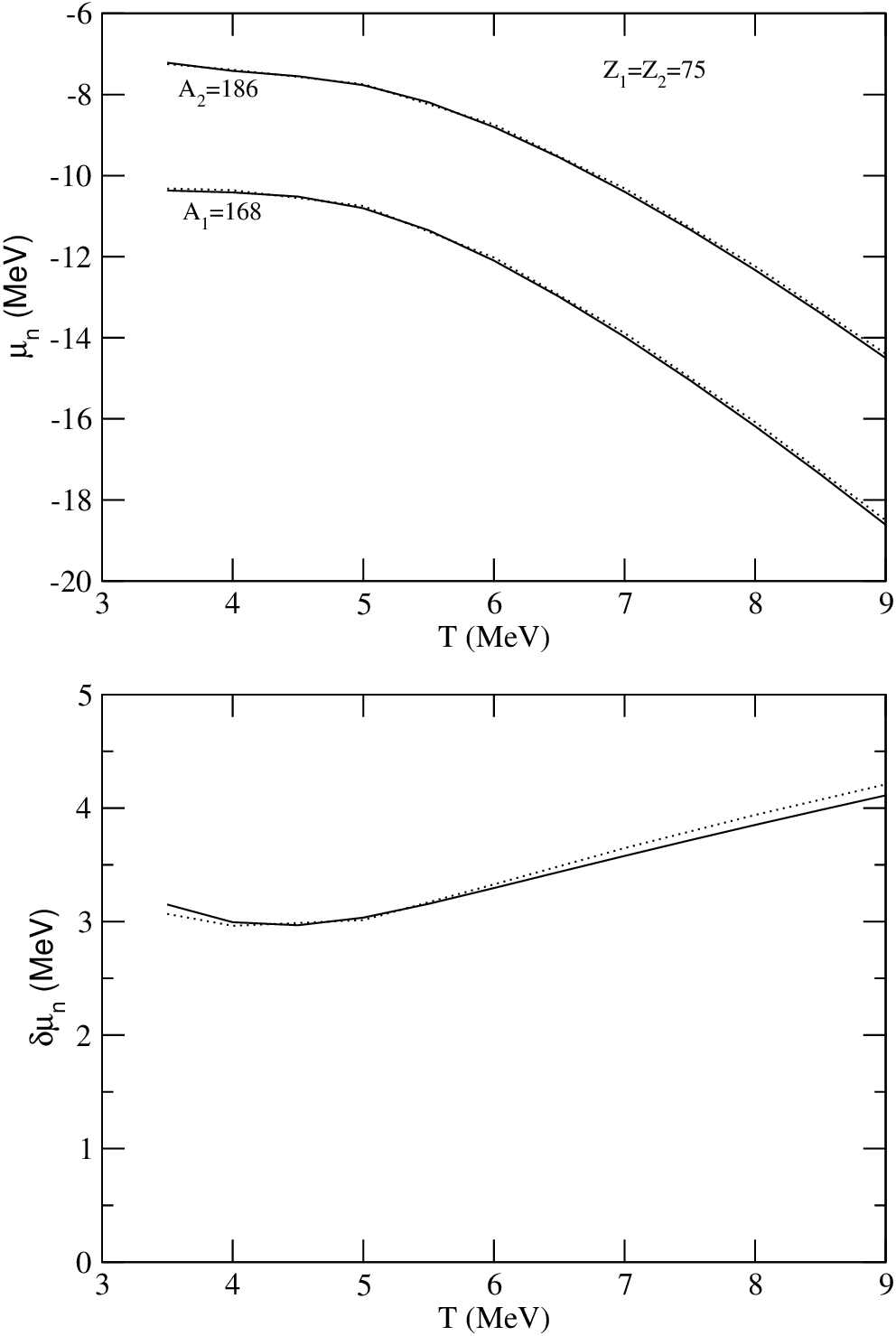}
\caption{ Chemical potentials $\mu_n$ (top panel) for systems $Z_1=75,
A_1=168$ and $Z_2=75, A_2=186$.  Canonical model (dashed curve) 
and grand canonical model (solid line) results are quite close.
The bottom panel shows results for $\delta\mu_n$ in a much more
enlarged scale.} 
\end{figure}

\begin{figure}
\includegraphics[width=5.5in,height=6.0in,clip]{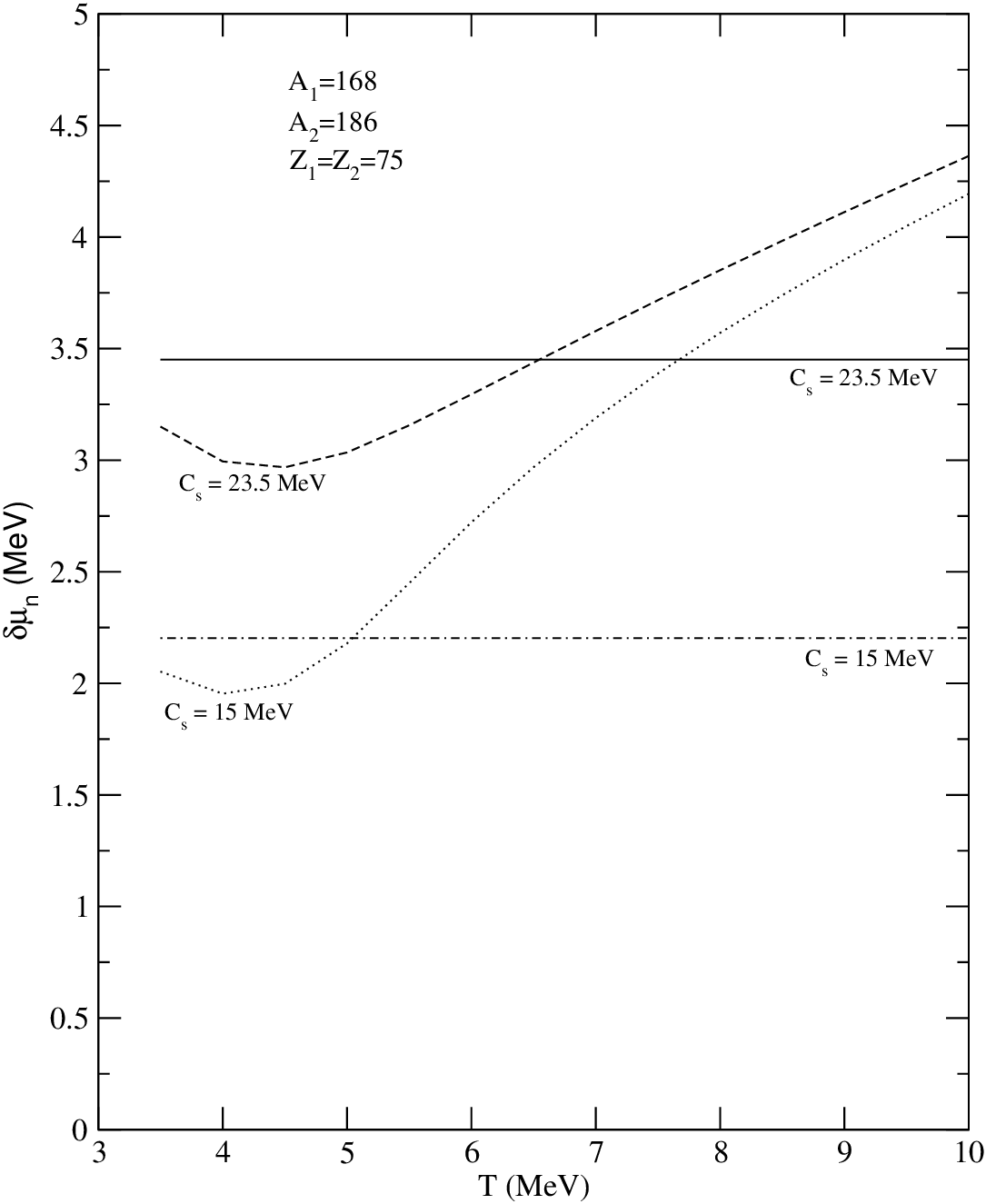}
\caption{ The difference $\delta\mu_n$ calculated from theory compared with
the simple prediction of eq.(4).  Since canonical and grand canonical
results for $\delta\mu_n$ are so close we just display the canonical model
results.  Two values of the symmetry energy coefficients were used.  Notice
that thermodynamic model predictions for $\delta\mu_n$ depend on the
temperature unlike the simple model prediction (eq.(4)).
In particular, irrespective of $C_s$, the theoretical value of $\delta\mu_n$
will approach the value $T\ln N_2/N_1$.}
\end{figure}

\begin{figure}
\includegraphics[width=5.5in,height=6.0in,clip]{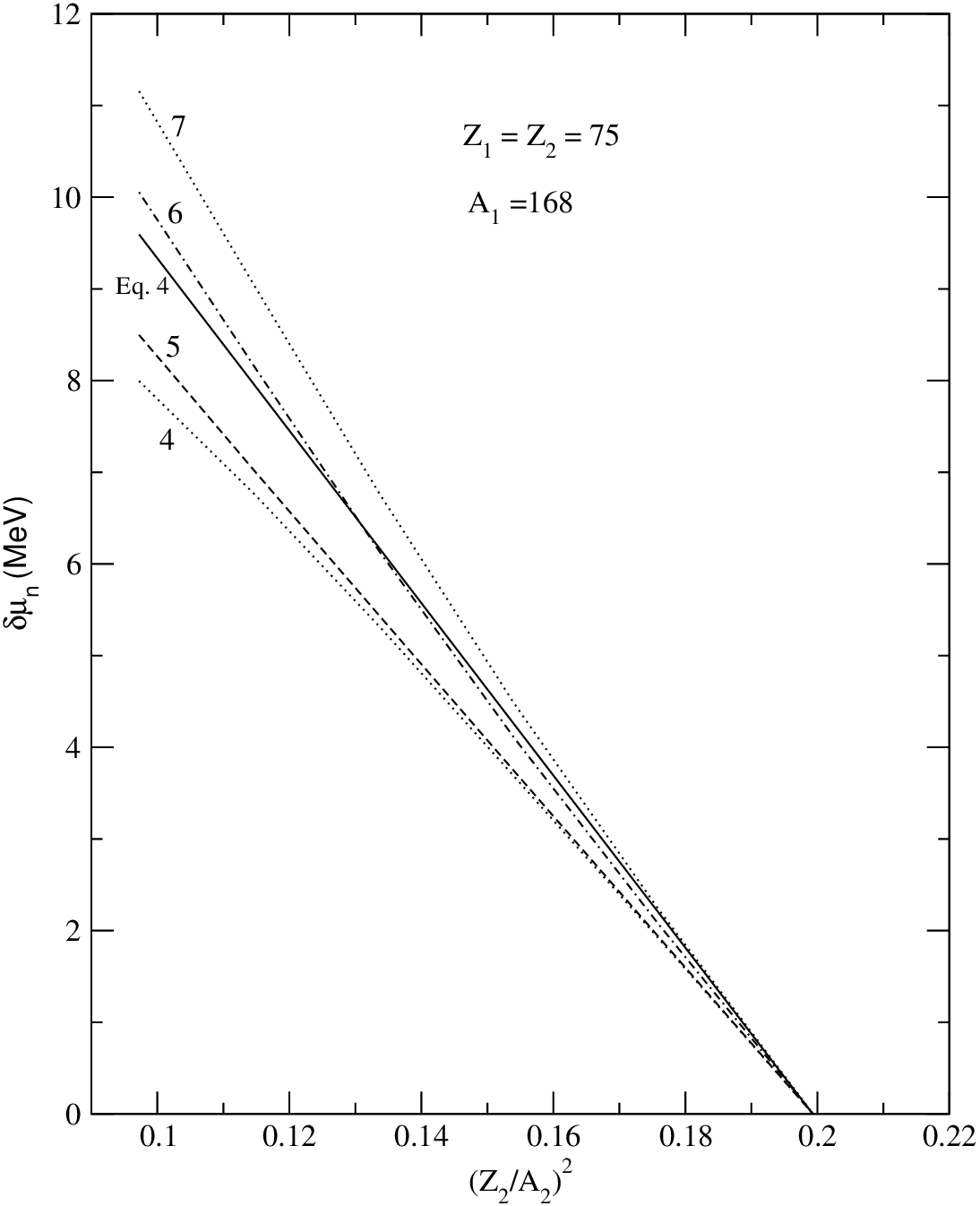}
\caption{ This figure tests the linearity of $\delta\mu_n$ against
$(Z_2/A_2)^2$ for fixed $(Z_1/A_1)$.  Canonical model calculations
are done for varying values of $Z_2/A_2$.  The calculations are done
with the value of $C_s$ at 23.5 MeV.  The solid line is a plot of eq.(4)
with $C_s$=23.5 MeV.  Notice that at 4 or 5 MeV interpreting the 
canonical model result using eq.(4) will lead to un underestimation
of the value of the underlying $C_s$ and an overestimation at 7 Mev
temperature.  Linearity in thermodynamic calculation is not perfect
but roughly correct in the temperature range displayed.}
\end{figure}

\begin{figure}
\includegraphics[width=5.5in,height=6.0in,clip]{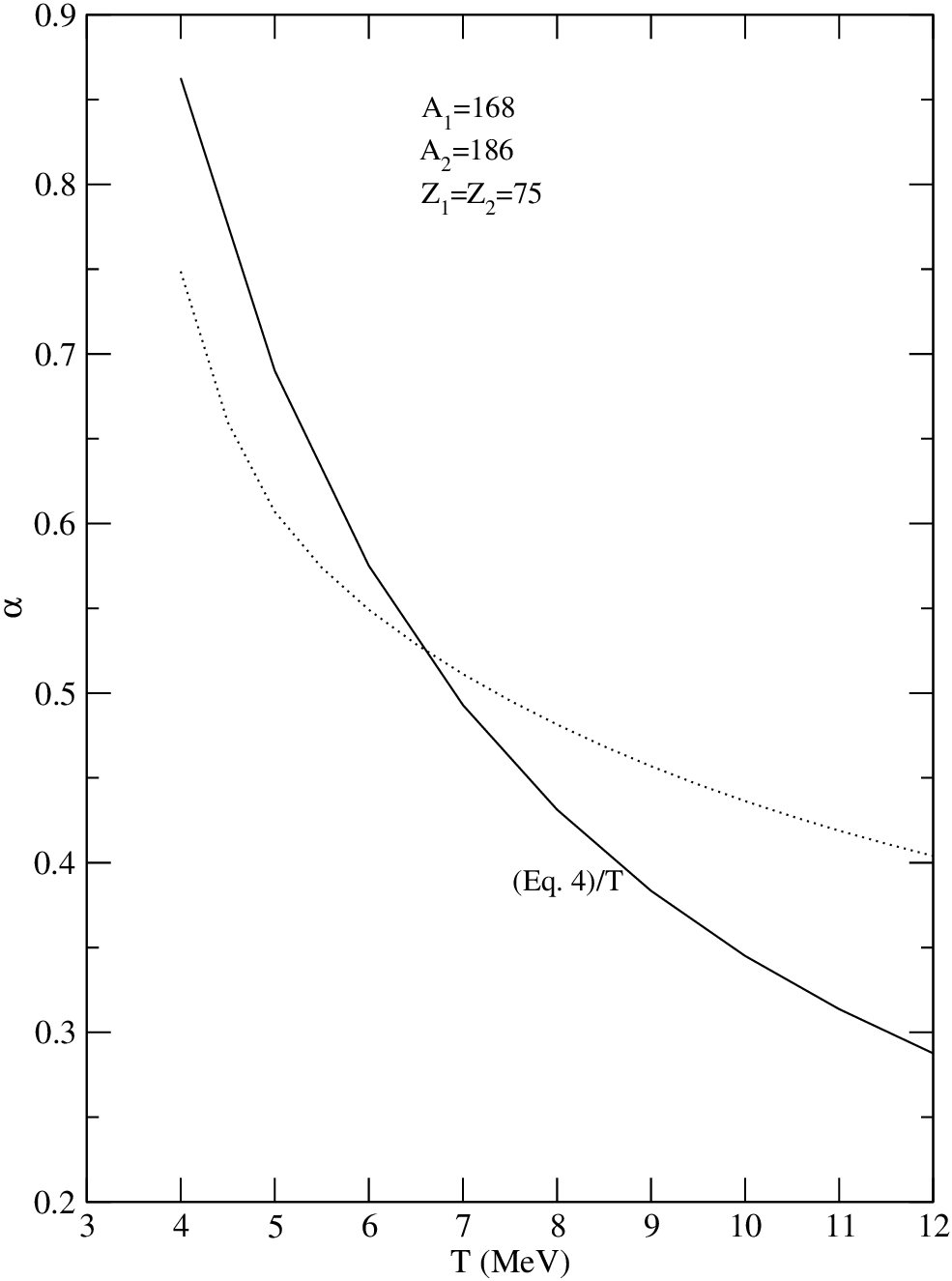}
\caption{The isoscaling parameter $\alpha=\delta\mu_n/T$ calculated
in the canonical model (the dotted curve) is campared with the standard
version $4(C_s/T)[(Z_1/A_1)^2-(Z_2/A_2)^2]$ (solid curve).
The thermodynamic model
predicts that $\alpha$ becomes constant asymptotically whereas
in standard parametrisation this would fall off like $1/T$.}
\end{figure}
\begin{figure}
\includegraphics[width=5.5in,height=6.0in,clip]{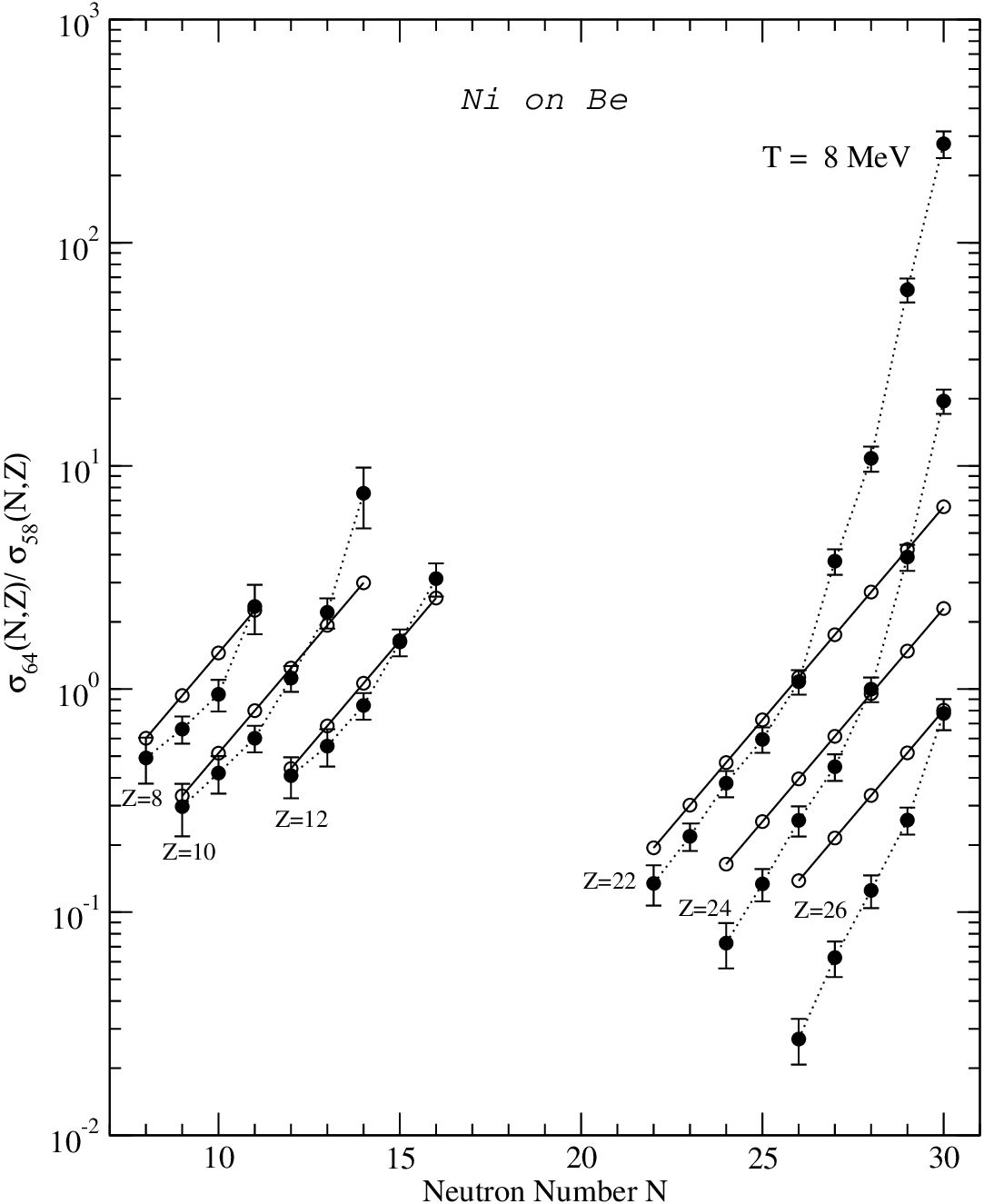}
\caption{Ratio of cross-sections of producing the nucleus (N,Z)
where reaction 1 is $^{58}$Ni on $^9$Be and reaction 2 is
$^{64}$Ni on $^9$Be, both at 140 MeV/n beam energy.  Experimental
data with error bars are compared with theoretical results from grand canonical ensemble ( hollow points).
Dotted lines are drawn through experimental points and solid lines 
through calculated points.  For calculation the dissociating
systems are taken to be Ni+Be.  The constant $C$ (eq.(42)) should be 
of the order 1 and in drawing the figure is taken to be 1.  The value
of the constant does not affect the slope of the log of the ratio.
Notice that for large ($N,Z$) the slopes are no longer those of a 
straight line and that ``average'' slopes for small ($N,Z$) and
large ($N,Z$) are very different in experiments but the same
in grand canonical calculation.}
\end{figure}

\begin{figure}
\includegraphics[width=5.5in,height=6.0in,clip]{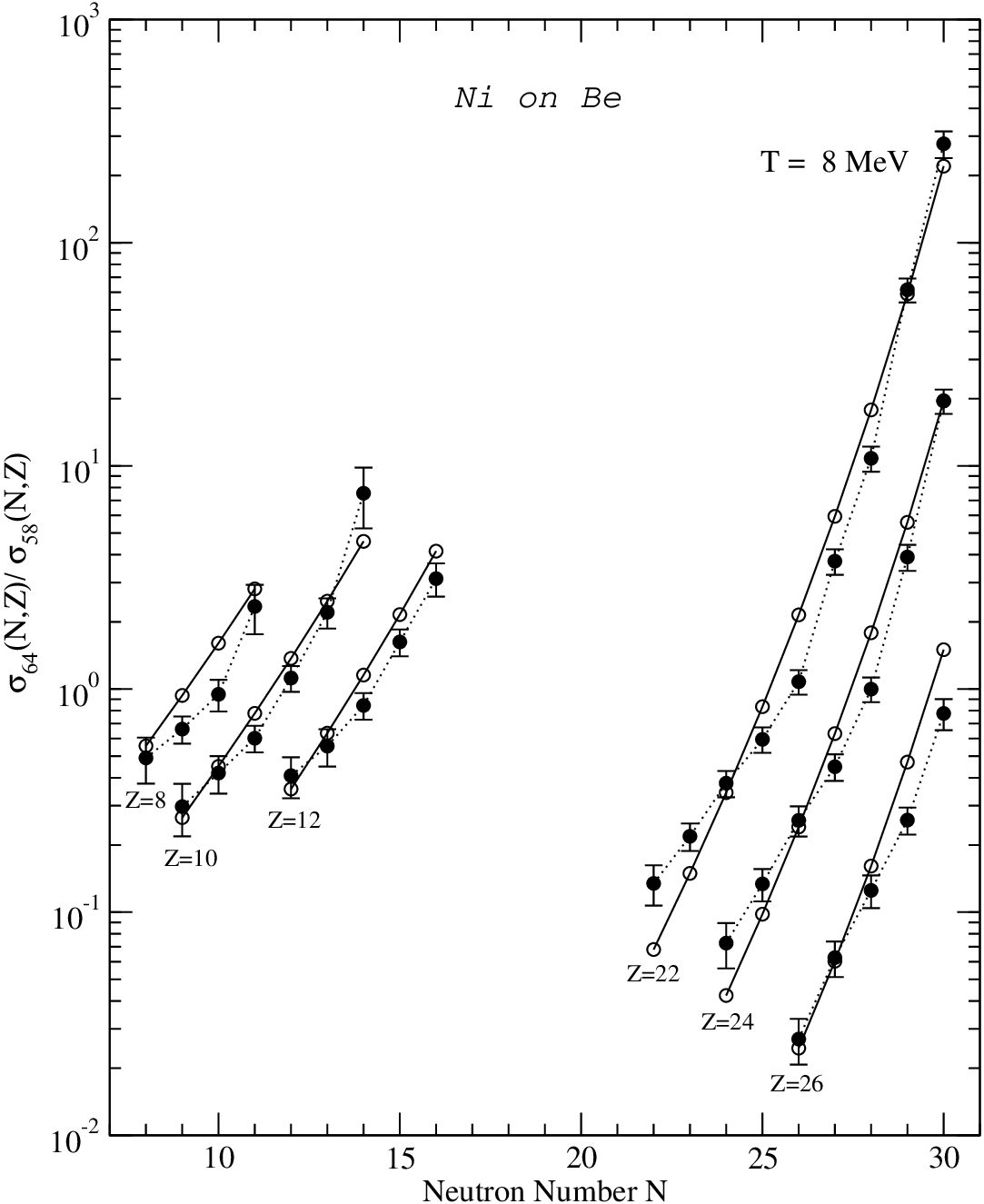}
\caption{The same as in Fig.6 except that the calculation is
done in the canonical ensmble.  Agreement with data is far
superior compared to that in Fig.6.}

\end{figure}
\begin{figure}
\includegraphics[width=5.5in,height=6.0in,clip]{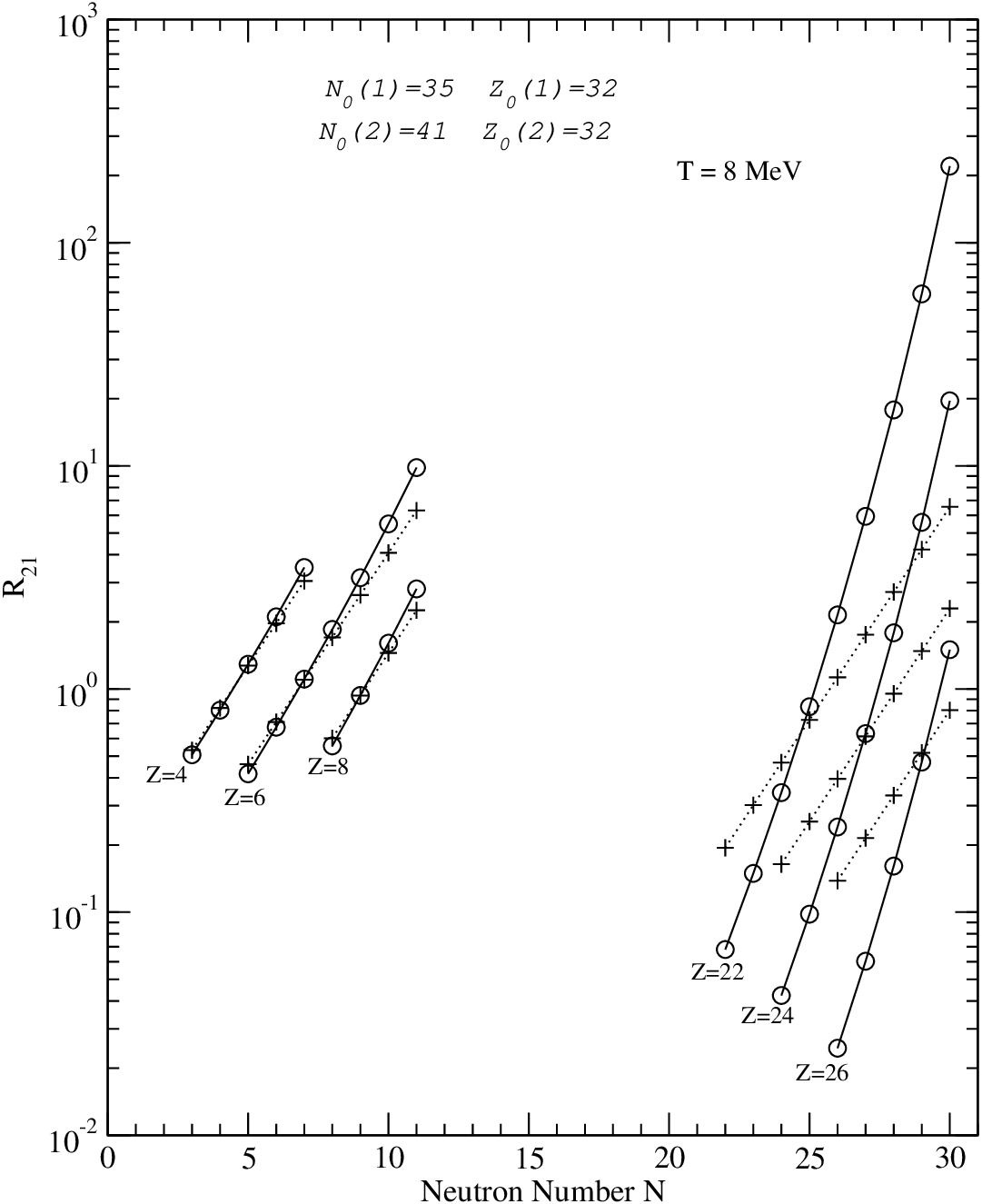}
:q
\caption{Comparisons of $R_{21}$ in canonical and grand canonical
models. The solid lines are through points computed in the
canonical model  and the dotted lines are through points calculated
in the grand canonical model.  The dissociating systems
are the same as in Figs. 6 and 7 and most of the points also
appear in those figures. Here  we
show results for lower $N,Z$ also to emphasize that
for low $N,Z$ agreement between canonical and grand canonical
is very good (and for these isoscaling works) 
but that they differ widely for large $N,Z$.
The grand canonical model should be discarded for large $N,Z$.}

\end{figure}

\begin{figure}
\includegraphics[width=5.5in,height=6.0in,clip]{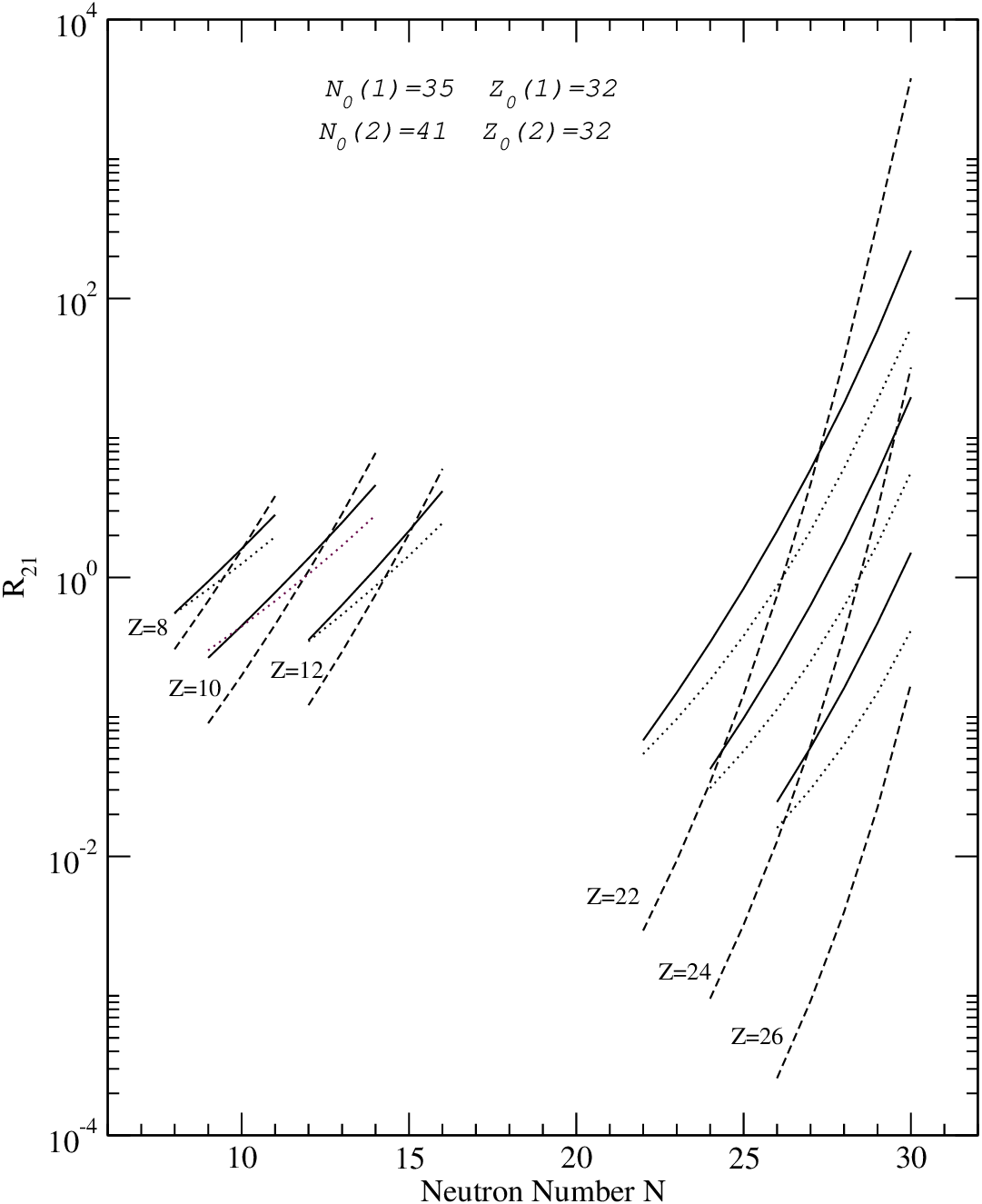}
\caption{The solid lines are canonical model results for $R_{21}$ 
at temperature 8
MeV (no effects of secondary decay).  This already fits the
data quite well (Fig.7).  For temperature 4.5 MeV 
we show two results.  The dashed lines are the direct calculations
(no corrections for decay).  The slopes are much higher than 
the 8 MeV curves.  The dotted curves
include a phenomenological correction (see section IX) due
to decay.  It brings down the slope and 
mimics the raw calculation at 8 MeV (solid line here)
and hence would fit the data reasonably well.}

\end{figure}

\end{document}